\documentclass[english,12pt]{article}

\usepackage{amsxtra,amssymb,amsthm,amsmath,latexsym}
\usepackage{booktabs}
\usepackage{multirow}
\usepackage{afterpage}
\usepackage{microtype}
\usepackage{calc,fourier}
\usepackage[T1]{fontenc}
\usepackage{hyperref}
\usepackage{float}
\usepackage[small]{caption}
\allowdisplaybreaks

\usepackage{setspace,cite}
\usepackage{babel}
\usepackage{verbatim}
\usepackage{epsf,graphicx}

\begin{document}
	
\newtheorem{thm}{Theorem}[]
\newtheorem{lem}[]{Lemma}
\newtheorem{defn}[]{Definition}
\numberwithin{equation}{section}
\newcommand{\ol}{\overline}
\newcommand{\ttbs}{\char'134}
\newcommand{\R}{{\mathbb R}}
\newcommand{\N}{{\mathbb N}}
\newcommand{\nb}{\nabla}
\newcommand{\la}{{\langle}}
\newcommand{\ra}{{\rangle}}
\newcommand{\dl}{{\delta}}
\newcommand{\bee}{\begin{equation*}}
\newcommand{\eee}{\end{equation*}}
\newcommand{\be}{\begin{equation}}
\newcommand{\ee}{\end{equation}}
\newcommand{\ba}{\begin{align}}
\newcommand{\ea}{\end{align}}
\newcommand{\pn}{\par\noindent}
\newcommand{\RRR}{\mathbb{R}^3}
%
\newcommand{\bZ}{{\mathbf Z}}
\newcommand{\bQ}{{\mathbf Q}}
\newcommand{\bR}{{\mathbf R}}
\newcommand{\cC}{{\mathcal C}}
%
\newcommand{\op}{{\rm \oplus}}
\newcommand{\id}{\rm id}
%
\newcommand{\eps}{\ensuremath{\epsilon}}
\newcommand{\ga}{\ensuremath{\gamma}}

\setlength{\textheight}{23cm}			
\setlength{\textwidth}{15cm}			
\setlength{\topmargin}{-0.5in}
\setlength{\oddsidemargin}{0.4in} 	
\setlength{\evensidemargin}{0.4in} 	

\title{Applications of potential theoretic mother bodies in Electrostatics}
\author{N. T. Tran\footnote{Mailing address:  Mathematics Department, 138 Cardwell Hall, Manhattan, KS 66506} \\
	\small Department of Mathematics\\
	\small Kansas State University, Manhattan, KS 66506-2602, USA\\
	\small \texttt{*nhantran@ksu.edu} 
}

\date{}

\maketitle
	
\begin{abstract}
Any polyhedron accommodates a type of potential theoretic skeleton called a mother body. The study of such mother bodies was originally from Mathematical Physics, initiated by Zidarov \cite{Zidarov:1990hr} and developed by Bj\"{o}rn Gustafson and Makoto Sakai \cite{GustafssonSakai:1999hr}. In this paper, we attempt to apply the brilliant idea of mother body to Electrostatics to compute the potentials of electric fields.
\end{abstract}



\section{Introduction} \label{sect:IntroIntro}

A mother body for a heavy body in geophysics is a concentrated mass distribution sitting inside an object (body), providing the same external gravitational field as the body. In this definition, a (heavy) body is a compact subset of $\bR^{n}$ provided with a mass distribution. The term "mother body" appears often in the geophysical literature. It was first rigorously defined by Bj\"{o}rn Gustafsson \cite{Gustafsson:1998hr} though the notion mother body dates back at least to the work of Bulgarian geophysicist Dimiter Zidarov \cite{Zidarov:1990hr}. Mother bodies are a powerful tool in geophysics since they provide a simple way to compute the gravitational field of objects. The problem of finding mother bodies is related to constructing a family of bodies that generate the same potential as a distributed mass. It was studied by many mathematicians and physicists like Zidarov \cite{Zidarov:1990hr}, Gustafsson  \cite{Gustafsson:1998hr,GustafssonSakai:1999hr}, Sakai \cite{GustafssonSakai:1999hr} and others.  

In image processing and computer vision fields, mother bodies or straight skeletons of two dimensional objects are defined as the locus of discrete points in raster environment that are located in the center of circles inscribed in the object that touch the object boundary in at least two different points \cite{FelkelObdrzalek}. The structure of straight skeletons is made up of straight line segments which are pieces of angular bisectors of polygon edges. The pattern recognition literature uses it heavily as a one-dimensional representation of a two dimensional object. 

The mathematical problem of constructing mother bodies is not always solvable, and the solution is not always unique \cite{SavinaSterninShatalov}. Indeed, there exist a number of bodies producing the same external gravitational field or external Newtonian potential in general \cite{SavinaSterninShatalov}. These bodies are called graviequivalent bodies or a family of graviequivalent bodies. Thus, it is essential to characterize each family by finding a mother body in it. Such an attempt was first done by Zidarov \cite{Zidarov:1990hr} in 1968. Later, many people have contributed different algorithms to solve this problem.

In this paper, we explore the mother bodies for convex polyhedra, assuming that any convex polyhedron preserves a unique mother body called a skeleton. The existence and uniqueness of mother body for convex polyhedra are proved by Gustafsson in paper \cite{Gustafsson:1998hr}. Furthermore, we attempt to apply the brilliant idea of using mother bodies to compute the Newtonian potential to Electrostatics.  Nevertheless, the computation is rather complicated since we must find a way to concentrate the electric charges in Electrostatics to the mother bodies of objects to ensure the potential produced by the mother bodies is the same as the one produced by the bodies. 

This paper is organized as follows: section \ref{sect:IntroPreli} introduces some basic notations and preliminaries relating to convex polyhedra and mother bodies. Section \ref{sect:MoBodyBody} and \ref{sect:MoBodyMoBody} discuss more about bodies and mother bodies. Section \ref{sect:MoBodyConvex} provides an important theorem on the existence and uniqueness of mother bodies for convex polyhedra stated in paper \cite{Gustafsson:1998hr}. In section \ref{sect:MoBodyExam} we apply the mother body method to Electrostatics to compute theoretic potentials. 

\section{Preliminaries} \label{sect:IntroPreli}
\subsection{Common terminology} \label{subsect:IntroTermi}
\begin{itemize}				
	\item \textbf{Hausdorff measure} 		: $\mathcal{H}^{n-1}$ denotes (n-1)-dimensional Hausdorff measures. $\mathcal{H}^{n-1}\left\lfloor\partial\Omega\right.$ denotes the (n-1)-dimensional Hausdorff measure restricted to $\partial\Omega$.
	
	\item \textbf{Lebesgue measure}  		: $\mathcal{L}^{n}$ denotes n-dimensional Lebesgue measures. $\mathcal{L}^{n}\left\lfloor\Omega\right.$ denotes the n-dimensional Lebesgue measure restricted to $\Omega$.
	
	\item \textbf{Newtonian kernel}	  	: $E(x)=\begin{cases}
	\frac{1}{2}x		&		(n = 1) \\
	-c_2 \text{log}|x|			&		(n = 2) \\
	c_n |x|^{2-n}		&		(n > 2)
	\end{cases}$ \\
	is the Newtonian kernel so that $-\Delta E = \delta$, the Diract measure at the origin.
	
	\item \textbf{Newtonian potential}  : $U^\mu = E * \mu$ , is the Newtonian potential of $\mu$, if $\mu$ is a distribution with 																	compact support in $\bR^n$, and $-\Delta U^\mu = \mu$.			
\end{itemize}

\subsection{Polyhedra} \label{subsect:IntroPoly}
\begin{defn}
	\textbf{(Convex polyhedron)} A \emph{convex polyhedron} in $\bR^n$ is a set of the form	
	\begin{equation}
	K = \bigcap_{i=1}^{n} H_i 	\label{equa:1}
	\end{equation}
	where $H_i$ are closed half-spaces in $\bR^n$, which satisfies: 
	\begin{align}
	&\text{int} \, K \neq \varnothing , \label{equa:2} \\
	&K \text{ is compact.} 			\label{equa:3}
	\end{align}
	A \emph{polyhedron} is a finite union (disjoint or not) of convex polyhedra as above. 
\end{defn}

\noindent\textbf{Remarks:}			
\begin{itemize}
	\item[(i)]   $P = \overline{\text{int} P}$ if $P$ is a polyhedron.
	\item[(ii)]  A polyhedron need not be connected.
	\item[(iii)] A convex polyhedron is a polyhedron which is convex as a set.
	\item[(iv)]  In the representation (\ref{equa:1}) of the convex polyhedron, the family \{$H_1$, \dots , $H_n$\} is unique provided n is taken to be \emph{minimal representation}.
	\item[(v)]   An equivalent definition of convex polyhedron is that it is a set which is the convex hull of finitely many points and having nonempty interior, see \cite{GustafssonSakai:1999hr}.
\end{itemize}

\begin{defn}
	\noindent \textbf{(Face of polyhedra)} For any set $P \subset \bR^n$, 
	\begin{align}
	\partial_{face}P = & \{x \in \bR^n : \text{ there exists } r>0 \nonumber \label{equa:4}\\
	& \text{and a closed half-space } H \subset \bR^n \text{ with } x \in \partial H \nonumber \\
	& \text{such that } P \cap B(x,r) = H \cap B(x,r) \}.
	\end{align}
	Then $\partial_{face}P$ is a relatively open subset of $\partial P$.\\
	A \emph{face} of a polyhedron $P$ is a connected component of $\partial_{face} P$.
\end{defn}

\subsection{Mother bodies} \label{subsect:IntroMbody}
		If $\Omega$ is a bounded domain in $\bR^n$ provided with a mass distribution $\rho_\Omega$ (e.g., Lebesgue measure restricted to $\Omega$), another mass distribution $\mu$ sitting in $\Omega$ and producing the same external Newtonian potential as $\rho_\Omega$ is called a \emph{mother body} of $\Omega$, provided it is maximally concentrated in mass distribution and its support has Lebesgue measure zero\cite{Gustafsson:1998hr}. 	
	
\section{Body} \label{sect:MoBodyBody}	
\begin{defn} \label{def:Body}
	A \emph{body} is a bounded domain $\Omega \subset \bR^n$ such that: 
	\begin{itemize}
		\item It is compact: $\Omega = \overline{\text{int} \, \Omega}$, 
		\item Its boundary has finite Hausdorff measure: $\mathcal{H}^{n-1}(\partial \, \Omega) < \infty $ 
		\item It is provided with an associated mass distribution $\rho = \rho_\Omega$.
	\end{itemize}
\end{defn}

 Frequently, the mass distribution in the domain is regarded as density one and outside is density zero. That means:
\begin{itemize}
	\item Inside $\Omega$: $\rho = \mathcal{L}^n \left\lfloor \Omega \right.$ , 
	\item On the boundary, $\partial \Omega$: $\rho = a \mathcal{H}^{n-1} \left\lfloor \partial H_j \right.$. 
\end{itemize}

 Therefore, given any two constant $a, b \geq 0$, with $a + b > 0$, we can associate with any $\Omega$ as above the mass distribution: 
\begin{equation} 
\rho_\Omega = a \mathcal{H}^{n-1} \left\lfloor \partial \Omega \right. + b \mathcal{L}^n \left\lfloor \Omega \right. 
\end{equation} 
Then, $\rho_\Omega$ is a positive Radon measure. We denote by $U^\Omega$ its Newtonian potential: 
\begin{equation} 
U^\Omega = U^{\rho \Omega} = E*\rho_\Omega 
\end{equation}
in which E is the Newtonian kernel, the Dirac measure at the origin.

\section{Mother body} \label{sect:MoBodyMoBody}
\begin{defn} \label{def:MoBody}
	Let $\Omega \subset \bR^n$ be a compact set satisfying $\Omega = \overline{\text{int} \, \Omega}$ and $U^\Omega$ be its Newtonian potential. $\Omega$ is regarded as a body with volume density one. A \emph{mother body} for $\Omega$ is a Radon measure $\mu$ satisfying these properties:	
	\begin{align}
	&\text{\textbullet\;\:} U^{\mu} = U^\Omega \text{ in } \bR^n \setminus \Omega,	\text{outside the body, the 																			potentials generated}  																	& \nonumber \\ 
	&\quad \text{ by the mother body and the body are the same,}										& \label{axiom:1} \\
	&\text{\textbullet\;\:} U^{\mu} \geq U^\Omega \text{ in } \bR^n, 								& \label{axiom:2} \\
	&\text{\textbullet\;\:} \mu \geq 0, 																						& \label{axiom:3} \\
	&\text{\textbullet\;\:} \text{supp} \, \mu \text{ has Lebesgue measure zero,} 					& \label{axiom:4} \\
	&\text{\textbullet\;\:} \text{For every } x \in \Omega \setminus \text{supp} \, \mu , \text{there exists a curve } 																			\gamma \text{ in } \bR^n \setminus \text{supp} \, \mu  				& \nonumber  \\ 
	&\quad \text{ joining } x \text { to some point in } \Omega^c.									& \label{axiom:5}
	\end{align}	
\end{defn}

\begin{figure}[htbp]
	\centering
	\includegraphics[width=0.25\textwidth]{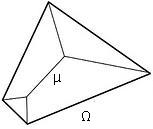}
	\caption{The mother body of a convex polyhedron}
	\label{fig:mbodyconvexpoly}
\end{figure}	

 From (\ref{axiom:1}) and (\ref{axiom:2}), it implies that: $\text{supp} \, \mu \subset \bar{\Omega}$, since the potential generated outside the body is positive but inside the body is zero. In order to find a measure $\mu$ that satisfies all of (\ref{axiom:1}) to (\ref{axiom:4}), we can just fill $\Omega$ with infinitely many disjoint balls until the remaining set has measure zero. Then, we can replace the mass distribution $\rho_\Omega$ by the sum of the appropriate point masses sitting in the center of these balls. Thus, we can rewrite the body and the mother body as:
\begin{align}
	&\Omega = \bigcup_{j=1}^\infty B(x_j,r_j) \cup \text{( Null set), where } B(x_j,r_j) \text{ are disjoint}, \\
	&\mu = a \mathcal{H}^{n-1} \left\lfloor \partial \Omega \right. + b \sum\limits_{j=1}^{\infty} \mathcal{L}^n (B(x_j,r_j))\delta_{x_j} 
\end{align}	
where $\delta_{x_j}$ denote the unit point mass at $x_j \in \bR^n$.

The five properties in definition \ref{def:MoBody} above are the basic axioms of mother bodies that are further discussed in paper \cite{Gustafsson:1998hr}. Note that since $\mu = \rho_\Omega$, the mother body is the concentrated mass distribution of the body, there exist many mother bodies satisfying axioms (\ref{axiom:1}) to (\ref{axiom:3}). Thus, a mother body for $\Omega$ should be one of them.  Nevertheless, there is neither existence nor uniqueness of mother bodies satisfying the five axioms in general; but under some special conditions such as in convex polyhedra case, the existence and uniqueness of solutions to the problem of finding mother bodies do hold. In the next section, we will investigate more about them.

\section{Mother bodies for convex polyhedra} \label{sect:MoBodyConvex}
 The following theorem is the most important result of studying mother bodies for convex polyhedra. It is stated in paper \cite{Gustafsson:1998hr}.
\begin{thm} \label{theorem:1}
	Let $\Omega \subset \bR^n$ be a convex bounded open polyhedron provided with a mass distribution $\rho_\Omega$ as in section \ref{sect:MoBodyBody}. Then there exists a measure $\mu$ satisfying axioms (\ref{axiom:1}) to (\ref{axiom:5}). Its support is contained in a finite union of hyperplanes and reaches $\partial \, \Omega$ only at corners and edges (not at faces), it has no mass on $\partial \, \Omega$, and $U^\mu$ is a Lipschitz continuous function. Moreover, $\mu$ is unique among all signed measures satisfying axioms (\ref{axiom:1}), (\ref{axiom:4}) and (\ref{axiom:5}).
\end{thm}
For the detailed proof, one can refer to paper \cite{Gustafsson:1998hr}. 
	
\section{Applications in Electrostatics}\label{sect:MoBodyExam}
In Physics, not every problem can be solved perfectly with a precise solution. Instead, we have to use approximation methods. Likewise for electric potentials, it is only for a few cases that we can compute the potentials exactly \cite{ClassicalMechanicsKibble}. In this section, we will use mother bodies to find the electric potentials of some basic bodies.

\subsection{Spherically uniform charge bodies}								
\begin{figure}[htbp]
	\centering
	\includegraphics[width = 2.5in]{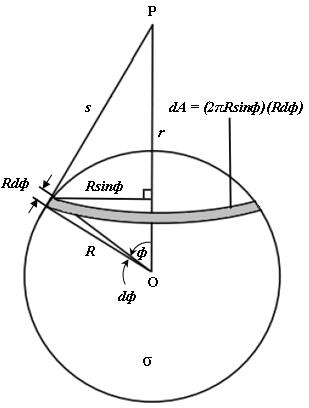}
	\caption{A spherical shell uniform charge body}
	\label{fig:ElectroSphericalShellCharge}
\end{figure}

Suppose we have a point P outside of a spherical shell S with radius $R$, area $A$ and uniform charge density $\sigma$ as showed in figure \ref{fig:ElectroSphericalShellCharge}. 

 Consider a ring on the shell, centered on the line from the center of the shell to P. Every particle on the ring has the same distance s to the point P. 

The electric potential of a particle $i$ on the ring with charge density $\sigma$ at P is:
\begin{equation}
V_i = \frac{1}{4 \pi \epsilon_0} \frac{\sigma}{s} \;
\end{equation}
in which $\epsilon_0$ is the permittivity of free space, $\epsilon_0 = 8.85 \times 10^{-12} \; \mathtt{C}^2 / N m^2$, and $s$ is the distance from the point P to the particle $i$. The electric potential of the whole ring at the point P is the sum of all potential $V_i$ at P. Let this potential sum be $\mathrm{d}V$. We have:
\begin{equation}
\mathrm{d}V = \sum_i{V_i} = \sum_{ring}{\frac{1}{4 \pi \epsilon_0} \frac{\sigma}{s}} = \frac{1}{4 \pi \epsilon_0 s} \sigma \mathrm{d}A \;,
\end{equation}
We need to find $\mathrm{d}A$, the area of one ring.
Let $\phi$ be the angle between two lines, the first one from the center of the sphere to the point P and the second one from the center of the sphere to the ring.\\
Then the radius of the ring is $R \, sin\phi$, the circumference of the ring is $2\pi R \, sin\phi$ and the width of the ring is $R \, d\phi$. Thus, the area of the ring is: 
\begin{equation}
\mathrm{d}A = 2\pi R^2 sin \phi \, \mathrm{d}\phi \;.
\end{equation}				
The total potential of the spherical shell S at the point P is the integral of $dV$ over the whole sphere as $\phi$ varies from 0 to $\pi$ and s varies from $r - R$ to $r + R$, where $r$ is the distance from the point P to the center of the spherical shell.
We need to rewrite $\phi$ in terms of $s$ to evaluate the integral. Using the Pythagorean theorem, we have:
\begin{equation}
s^2 = (r - R cos \phi)^2 + (R sin \phi)^2 = r^2 - 2 r R cos \phi + R^2 \;.
\end{equation}
Taking differentials of both sides:
\begin{equation}
2 s \, \mathrm{d}s = 2 r R \, sin \phi \, \mathrm{d}\phi \quad\rightarrow\quad sin \phi \, \mathrm{d}\phi = \frac{s \, \mathrm{d}s}{r R} \;.
\end{equation}
Hence, we have: 
\begin{equation}
\mathrm{d}A = \frac{2\pi R^2 s \mathrm{d}s}{rR} = \frac{2\pi R s \mathrm{d}s}{r} \;.
\end{equation}		
Substituting this into $dV$:
\begin{equation}
\mathrm{d}V = \frac{1}{4 \pi \epsilon_0 s} \sigma \mathrm{d}A = \frac{\sigma}{4 \pi \epsilon_0 s} \frac{2 \pi R s \mathrm{d}s}{r} = \frac{\sigma R}{2 \epsilon_0 r} \mathrm{d}s \;.
\end{equation}	
Integrating $\mathrm{d}V$, we have:
\begin{equation}
V = \frac{\sigma R}{2 \epsilon_0 r} \int_{r - R}^{r + R}\mathrm{d}s = \frac{\sigma R}{2 \epsilon_0 r}[(r + R) - (r - R)] = \frac{\sigma R^2}{\epsilon_0 r} \;.
\end{equation} 
Since the total charge of the shell is $q = 4 \pi R^2 \sigma$, we have: $\sigma = q/4 \pi R^2$. \\
Thus, we can rewrite $V$:
\begin{equation}
V = \frac{1}{4 \pi \epsilon_0} \frac{q}{r} \;.
\end{equation}
This is the electric potential of a point charge $q$ at the distance $r$, i.e. the potential of the spherical shell S (body) at any distance $r$ is the same as that of its mother body at the same distance.

\subsection{Cylindrically uniform charge bodies}
\begin{figure}[htbp]
	\centering
	\includegraphics[width = 3.5in]{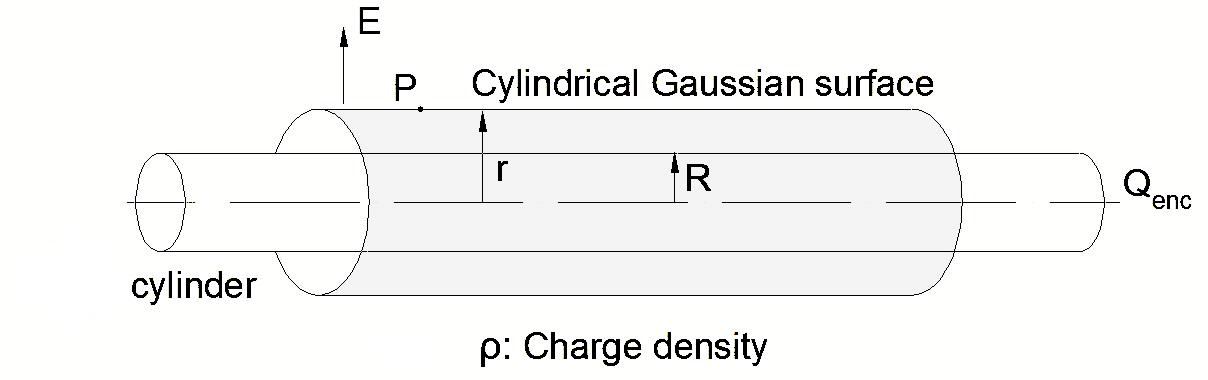}
	\caption{A cylindrical uniform charge body}
	\label{fig:ElectroCylindricalCharge}
\end{figure}

 Let P be a point outside a cylindrical solid body with radius $R$, length $L$, and uniform charge density $\rho$ as in figure \ref{fig:ElectroCylindricalCharge}. The distance from P to the center axis of the cylinder is $r$. 

We will create a Gaussian surface around the cylinder as in figure \ref{fig:ElectroCylindricalCharge}. By Gauss's law, the total flux of the electric field $E$ over this Gaussian surface is:
\begin{equation}
\int_{surface} E_{cylinder} \cdot \mathrm{d}\mathbf{A} = \frac{Q_{enc}}{\epsilon_0}  \;, \label{ExElecCylinderGauss}
\end{equation}
where $Q_{enc}$ is the total charge enclosed within the surface $S$ and $\mathrm{d}\mathbf{A}$ is the area of an infinitesimal piece of the surface $S$.

Since the electric field is the same everywhere by symmetry, we can move $E_{cylinder}$ out of the integral. Furthermore, since $E_{cylinder}$ points radially outward, as does $\mathrm{d}A$, we can drop the dot product and deal with only the magnitudes:
\begin{align}
\int_{surface} E_{cylinder} \cdot \mathrm{d}\mathbf{A} &= \int_{surface} |E_{cylinder}| \mathrm{d}A \\
																					&= |E_{cylinder}| \int_{surface} \mathrm{d}A \\
																					&= |E_{cylinder}| 2 \pi r L \; ,
\end{align}
in which $L$ is the length of the cylinder and $r$ is the distance from the center of the cylinder to the cylindrical Gaussian surface.

The total charge $Q_{enc} = \pi R^2 L \rho$. Substituting all these into \ref{ExElecCylinderGauss}, we have:
\begin{align}
\int_{surface} E_{cylinder} \cdot \mathrm{d}\mathbf{A} = \frac{Q_{enc}}{\epsilon_0} &\quad\Leftrightarrow\quad |E_{cylinder}| 2 \pi r L = \frac{\pi R^2 L \rho}{\epsilon_0} \\ 																																&\quad\Leftrightarrow\quad |E_{cylinder}| 2 r = \frac{R^2 \rho}{\epsilon_0} \;.
\end{align}
Hence:
\begin{equation}
|E_{cylinder}| = \frac{R^2 \rho}{2 \epsilon_0 r} \; ,
\end{equation}
or:
\begin{equation}
E_{cylinder} = \frac{R^2 \rho}{2 \epsilon_0 r}\hat{r} \;,
\end{equation}
where $\hat{r} $ is the unit vector pointing in the direction of $E$.

If we choose the reference point at a distance $a$, and since $E_{cylinder}$ points outward, the potential is:
\begin{equation}
V_{cylinder} = - \int_{a}^{r} E_{cylinder} \cdot \mathrm{d}\mathbf{l} = - \int_{a}^{r} \frac{R^2 \rho}{2 \epsilon_0 r}\mathrm{d}r = - \left. \frac{R^2 \rho}{2 \epsilon_0} \text{ln}|r| \right|_{a}^{r} = -\frac{R^2 \rho}{2 \epsilon_0} \text{ln}\left(\frac{r}{a}\right) \;, \label{ExElecCylinderPotential}
\end{equation}
where $\mathrm{d}\mathbf{l} = \hat{r}\mathrm{d}r$.

Now, let us consider the mother body of this cylinder. Its mother body is the symmetric axis. Since the mother body has the same total charge with the body, $Q_{enc} = \pi R^2 L \rho$, its charge density is $\lambda = \pi R^2 \rho$. 

\begin{figure}[htbp]
	\centering
	\includegraphics[width = 3.5in]{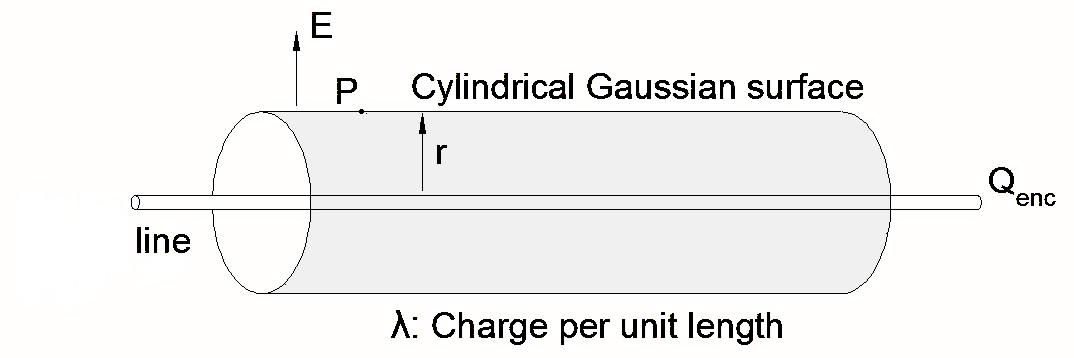}
	\caption{A line charge}
	\label{fig:ElectroLineCharge}
\end{figure}

We will also create a Gaussian surface around this line as in figure \ref{fig:ElectroLineCharge}. By Gauss's law, the total flux of the electric field over this Gaussian surface is:
\begin{equation}
\int_{surface} E_{line} \cdot \mathrm{d}\mathbf{A} = \frac{\lambda L}{\epsilon_0} \; , \label{ExElecCylinderMomGauss}
\end{equation}
in which $L$ is the length of the line.\\
By the same argument as above, \ref{ExElecCylinderMomGauss} becomes:
\begin{equation} 
|E_{line}| 2 \pi r L = \frac{\lambda L}{\epsilon_0} \quad\Leftrightarrow\quad |E_{line}| 2 \pi r = \frac{\lambda}{\epsilon_0} \;.
\end{equation}
Hence:
\begin{equation}
|E_{line}| = \frac{\lambda}{2 \epsilon_0 \pi r} \; ,
\end{equation}
or:
\begin{equation}
E_{line} = \frac{\lambda}{2 \epsilon_0 \pi r}\hat{r} \;.
\end{equation}
If we also choose the reference point at a distance $a$, and since $E_{line}$ points outward, the potential is:
\begin{equation}
V_{line} = - \int_{a}^{r} E_{line} \cdot \mathrm{d}\mathbf{l} = - \int_{a}^{r} \frac{\lambda}{2 \epsilon_0 \pi r}\mathrm{d}r = - \left. \frac{\lambda}{2 \epsilon_0 \pi} \text{ln}|r| \right|_{a}^{r} = -\frac{\lambda}{2 \epsilon_0 \pi} \text{ln}\left(\frac{r}{a}\right) \;.
\end{equation}
Substituting $\lambda = \pi R^2 \rho$ into $V_{line}$, we have:
\begin{equation}
V_{line} = -\frac{\pi R^2 \rho}{2 \epsilon_0 \pi} \text{ln}\left(\frac{r}{a}\right) = -\frac{R^2 \rho}{2 \epsilon_0} \text{ln}\left(\frac{r}{a}\right) \;. \label{ExElecCylinderMomPotential}
\end{equation}
Obviously, the potential generated by the mother body is the same as one generated by the cylinder if we compare \ref{ExElecCylinderMomPotential} and \ref{ExElecCylinderPotential}.																					

\subsection{Conical uniform charge bodies}
\begin{figure}[H]
	\centering
	\includegraphics[width = 1.5in]{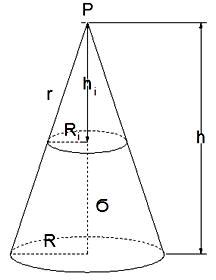}
	\caption{A conical uniform charge body}
	\label{fig:ElectroConicalCharge}
\end{figure}

 Let P be the point sitting on the vertex of a conical surface (an empty cone) with radius $R$, height $h$ and uniform charge density $\sigma$ as in figure \ref{fig:ElectroConicalCharge}. 

If we divide the cone into rings, the electric potential of an element on one ring at the point P is:
\begin{equation}
V_i = \frac{1}{4 \pi \epsilon_0} \frac{\sigma}{r} \; ,
\end{equation}
in which $r$ is the distance from P to an element on the ring. Since P is on the axis of symmetry, $r$ is the same for every element on the ring, and $r = \sqrt{R_i^2 + h_i^2}$ where $h_i$ is the distance from P to the center of the ring and $R_i$ is the radius of the ring. The circumference of the ring is: $2 \pi R_i$.\\
Therefore, the electric potential of each ring to the point P is:
\begin{equation}
V_{ring} = \int V_i = \int_{0}^{2 \pi R_i}{\frac{1}{4 \pi \epsilon_0} \frac{\sigma}{r}} \mathrm{d}l = \frac{1}{4 \pi \epsilon_0} \frac{\sigma 2 \pi R_i}{r} = \frac{\sigma R_i}{2 \epsilon_0 r} \;.
\end{equation}											
In order to find the electric potential of the whole cone to the point P, we need to take the integral:
\begin{equation}
V_{cone} = \int V_{ring} = \int_{0}^{\sqrt{R^2 + h^2}} \frac{\sigma R_i}{2 \epsilon_0 r} \mathrm{d}r \;.				
\end{equation}
in which $\sqrt{R^2 + h^2}$ is the slant length of the cone. \\
Since we have:
\begin{equation}
\frac{R_i}{r} = \frac{R}{\sqrt{R^2 + h^2}} \; ,
\end{equation}
substituting this into $V_{cone}$, we have:								
\begin{align}
V_{cone} &= \int_{0}^{\sqrt{R^2 + h^2}} \frac{\sigma R_i}{2 \epsilon_0 r} \mathrm{d}r \nonumber\\											  									 &= \int_{0}^{\sqrt{R^2 + h^2}} \frac{\sigma R}{2 \epsilon_0 \sqrt{R^2 + h^2}} \mathrm{d}r \nonumber\\
&= \left. \frac{\sigma R}{2 \epsilon_0 \sqrt{R^2 + h^2}} r \right|_{0}^{\sqrt{R^2 + h^2}} \nonumber\\	
&= \frac{\sigma R}{2 \epsilon_0} 			\;.										\label{ElecExCone1}
\end{align}																
We can rewrite this formula in terms of the total charge $q$ by a small replacement:
\begin{equation}
q = \pi R \sqrt{R^2 + h^2} \sigma \quad\rightarrow\quad \sigma = \frac{q}{\pi R \sqrt{R^2 + h^2}} \;. \label{ElecExConeChargeDensity}
\end{equation}
Substituting this into $V_{cone}$:
\begin{equation}
V_{cone} = \frac{R}{2 \epsilon_0} \frac{q}{\pi R \sqrt{R^2 + h^2}} = \frac{q}{2 \epsilon_0 \pi \sqrt{R^2 + h^2}} \;. \label{ElecExCone2}
\end{equation}

Now, let us consider the mother body of the cone. Its mother body is the symmetric axis connecting the vertex of the cone with the center of the bottom circle. Since the mother body has the same total charge with the body, $q$, its charge density function is $q_i = 2 \pi R_i \sigma$. \\
The electric potential of this mother body on the point P is:
\begin{equation}
V_{line} = \frac{1}{4 \pi \epsilon_0} \int_{0}^{h} \dfrac{q_i}{h_i} \mathrm{d}h_i = \frac{1}{4 \pi \epsilon_0} \int_{0}^{h} \dfrac{2 \pi R_i \sigma}{h_i} \mathrm{d}h_i \;.
\end{equation}
Since we have:
\begin{equation}
\frac{R_i}{h_i} = \frac{R}{h} \; ,
\end{equation}
substituting this into $V_{line}$: 				
\begin{equation}
V_{line} = \frac{1}{4 \pi \epsilon_0} \int_{0}^{h} \dfrac{2 \pi R \sigma}{h} \mathrm{d}h_i = \left. \frac{R \sigma}{2 \epsilon_0 h} h_i \right|_{0}^{h} = \frac{R \sigma}{2 \epsilon_0 h} h = \frac{\sigma R}{2 \epsilon_0} \;. \label{ElecExConeMom1}
\end{equation}		
Replacing $\sigma$ by using \ref{ElecExConeChargeDensity}, we have:
\begin{equation}
V_{line} = \frac{q}{2 \epsilon_0 \pi \sqrt{R^2 + h^2}} \;. \label{ElecExConeMom2}
\end{equation}	
Clearly, the potential generated by the mother body is the same as one generated by the cone if we compare \ref{ElecExConeMom2} and \ref{ElecExCone2}.

\section{Conclusions}\label{sect:Conclusion}
Now we know how to apply the idea of using mother bodies to compute the Newtonian potential to Electrostatics.  Moreover, we can see that the potential computation is much easier if we use mother bodies rather than bodies in general. However, it is easy only for simple symmetric bodies. For complicated bodies, e.g. asymmetric ones, the computation is very difficult, sometimes impossible to do in closed form, since we have to deal with integrals on mother bodies. 

Another problem is how to find an accurate distribution function for the potential of a mother body, e.g. charge distribution function for electric potential. Consider the case when an object is a uniformly charged square plate and its mother body is the diagonals. In this case, the charge density for the body is uniform, yet the charge density for its mother body is not. For the diagonals, the charge density is not a uniformly distributed function. Thus, the way we formulate the distribution function will affect the precision of the result of potential computation for the mother body. This problem is still open.


\begin{thebibliography}{1}
	
\bibitem{ClassicalMechanicsKibble}
{Kibble, T. W. B.}
\newblock {\em {Classical Mechanics}}.
\newblock John Wiley and Sons, New York, 2 edition, 1973.

\bibitem{Zidarov:1990hr}
{Zidarov, D.}
\newblock {\em On Solution of Some Inverse Problems for Potential Fields and
	Its Application to Questions in Geophysics.}
\newblock (Sofia: Publ. House of Bulg. Acad. of Sci.), 1990.

\bibitem{FelkelObdrzalek}
{Felkel, P. and Obdr\v{z}\'{a}lek, S.}
\newblock Straight skeleton implementation.
\newblock In {\em Proceedings of Spring Conference on Computer Graphics}, pages
210--218, Budmerice, Slovakia, 1998.

\bibitem{Gustafsson:1998hr}
{Gustafsson, B.}
\newblock On mother bodies of convex polyhedra.
\newblock {\em SIAM J. Math. Anal}, 29(5):1106--1117, 1998.

\bibitem{GustafssonSakai:1999hr}
{Gustafsson, B. and Sakai, M.}
\newblock On potential theoretic skeletons of polyhedra.
\newblock {\em Geometriae Dedicata}, 76:1--30, 1999.

\bibitem{SavinaSterninShatalov}
{Savina, T. V., Sternin, B. Yu. and Shatalov, V. E.}
\newblock On a minimal element for a family of bodies producing the same
external gravitational field.
\newblock {\em Applicable Analysis}, pages 649--668, 2005.


	
\end{thebibliography}

\end{document}